\documentclass[a4paper,10pt]{article}
\usepackage[margin=1.0in]{geometry}

\usepackage{amssymb,amsfonts,amsmath,epsfig,float,graphicx}

\usepackage {url}
\usepackage{accents}

\newtheorem{theorem}{Theorem}[section]
\newtheorem{lemma}[theorem]{Lemma}

\newcommand{\qed}{\nobreak \ifvmode \relax \else
      \ifdim\lastskip<1.5em \hskip-\lastskip
      \hskip1.5em plus0em minus0.5em \fi \nobreak
      \vrule height0.75em width0.5em depth0.25em\fi}

\def\upd{{\rm d}}
\newcommand{\be}{\begin{eqnarray}}
\newcommand{\ee}{\end{eqnarray}}

\def\Ro{{\mathbb R}}

\def\AA{{\bf A}}
\newcommand{\Abf}{{\bf A}}

\newcommand{\Ebf}{{\bf E}}

\def\kk{{\bf k}}

\def\xx{{\bf x}}
\def\yy{{\bf y}}

\def\uc{q}

\def\lu{{l}}
\def\mstar{\kappa}

\def\psicha{\hat\psi^{\mbox{\tiny a}}}

\renewcommand{\Re}{\,{\rm Re}\,}

\def\Hph{\hat H^{{\mbox{\tiny ph}}}}
\def\HI{\hat H^{{\mbox{\tiny I}}}}

\def\Hel{\hat H^{{\mbox{\tiny el}}}}

\def\ah{\hat a_{\mbox{\tiny H}}}
\def\av{\hat a_{\mbox{\tiny V}}}

\def\rhoel{\rho^{\mbox{\tiny el}}}

\def\Zph{Z^{\mbox{\tiny ph}}}
\def\Zel{Z^{\mbox{\tiny el}}}

\def\kkph{{\kk^{\mbox{\tiny ph}}}}

\def\kph{k^{\mbox{\tiny ph}}}

\begin{document}
\title{Reducible Quantum Electrodynamics.
III. The emergence of the Coulomb forces}
\author{Jan Naudts}

\maketitle

\begin{abstract}
The assumption is made that only transversely polarized photons are needed for a
correct description of Quantum Electrodynamics. A simple mathematical transformation
is used to introduce new field operators which satisfy the full Maxwell equations.
In particular, they reproduce Coulomb forces between different regions of the
charge field. The analogy with the polaron problem can give some insight in the
physics underlying the transformation. In this context it is shown that the 
interaction of the electron field with a transversely polarized photon field
can form bound states. The binding energy peaks for long wavelength photons.
\end{abstract}

\section{Introduction}

This is the third in a series of papers on reducible quantum electrodynamics (rQED).
In the first paper \cite {NJ15a},  denoted (I) hereafter, the free electromagnetic
field is studied. Free electron fields are the topic of the second paper \cite {NJ15b},
denoted (II) hereafter. The next step is to introduce interactions between free fields.
In standard QED the interaction Hamiltonian involves additional states of the 
electromagnetic field which do not appear in the description of free fields.
Associated with them are so-called longitudinal and scalar photons.
Longitudinal and scalar photons do not appear in the present work.
The obvious question is then how to quantize electrostatic fields.
This is the topic of the present paper.

\bigskip
Michael Creutz \cite{CM79} showed that a simple mathematical transformation can remove
the static fields produced by electric charges. In the next
section this idea in used in the opposite direction.
It results in two different pictures of the same physics, a Heisenberg picture
of interacting electron and photon fields, in absence of electrostatic fields,
and an {\em emergent picture} in which the field operators satisfy the full Maxwell
equations. The existence of this mathematical transformation suggests that the Coulomb forces
are emergent forces, much in the same sense as the recent claims \cite {VEP11,VEP16}
that gravity forces are emergent forces.

\bigskip
The remainder of the paper is an attempt to understand the physics
behind this double description of electromagnetic fields.
In Solid State Physics the {\sl polaron} is a bound state of an electron and quantized
lattice vibrations. An extensive body of knowledge about polarons exists --- see
for instance \cite{DA09}.
The Fr\"olich Hamiltonian \cite {FH54}, which is used to describe polarons,
is very similar to the Hamiltonian of QED.
Hence, by analogy one expects that free electron fields can bind with the photon field.

\bigskip
Section \ref {sect:bound} proves that transversely
polarized photons can indeed lower the total energy of an electron field.
Such a dressed electron field can be compared with a polaron.
It is known that polarons can attract each other.
This raises the question whether the Coulomb forces between different parts of the
electron field can be explained as effects due to the dressing with photons.
The final section discusses this point.

%%%%%%%%%%%%%%%%%%%%%%%%%%%%%%%%%%%%
%%%%%%%%%%%%%%%%%%%%%%%%%%%%%%%%%%%%
%%%%%%%%%%%%%%%%%%%%%%%%%%%%%%%%%%%%
%%%%%%%%%%%%%%%%%%%%%%%%%%%%%%%%%%%%
%%%%%%%%%%%%%%%%%%%%%%%%%%%%%%%%%%%%
%%%%%%%%%%%%%%%%%%%%%%%%%%%%%%%%%%%%
\section{Gauss' law}

\subsection{Hamiltonian}

The use of the temporal gauge is obvious because of the assumption that
longitudinal and scalar photons do not exist. In fact, what is used is sometimes called
the transverse gauge. It is the combination of the Coulomb gauge ($\nabla\cdot\AA=0$)
with the absence of any charges. The vector potential operator
$\hat A^0$ vanishes identically. The three operators $\hat A_\alpha$ are not independent.
They are defined by (see (I))
\be
\hat A_\alpha(x)&=&\frac 1{2}\lambda\varepsilon^{(H)}_\alpha(\kkph)
\left[e^{-i\kph_\nu x^\nu}\ah+e^{i\kph_\nu x^\nu}\ah^\dagger\right]
+\frac 1{2}\lambda\varepsilon^{(V)}_\alpha(\kkph)
\left[e^{-i\kph_\nu x^\nu}\av+e^{i\kph_\nu x^\nu}\av^\dagger\right],
\label{em:potop}
\ee
with polarization vectors $\varepsilon^{(H)}_\alpha(\kkph)$ and $\varepsilon^{(V)}_\alpha(\kkph)$.
The wave vector of the photon field is denoted $\kkph$.
The operators $\ah$ and $\av$ are the annihilation operators of horizontally,
respectively vertically polarized photons.
The constant $\lambda$ is there for dimensional reasons.
A drawback of the temporal gauge is the loss of manifest Lorentz covariance.

\bigskip
The Dirac field operators are given by (see (II))
\be
\hat\psi_{r,\kk}(x)
&=&
\sum_{s=1,2}u_r^{(s)}(\kk)\hat\phi_{s,\kk}^{(+)}(x)
+\sum_{s=3,4}v_r^{(s)}(\kk)\hat\phi_{s,\kk}^{(-)}(x).
\label{electron:defpsi}
\ee
They are used to define the electric current operators
\be
[\hat j^\mu(x)\psi]_{\kk}
&=&\frac {\uc c}{(2\pi)^3}\int\upd\kk'\, \hat J^\mu_{\kk,\kk'}(x)\psi_{\kk'}.
\label{electron:jdef}
\ee
Here, the operators $ \hat J^\mu(x)$ are defined by
\be
\hat J^\mu_{\kk,\kk'}(x)
&=&\frac {1}2
\sum_{r,r'}\gamma^\mu_{r,r'}\psicha_{r,\kk}(x)\hat\psi_{r',\kk'}(x)
-\frac {1}2\sum_{r,r'}\gamma^\mu_{r',r}\hat\psi_{r,\kk}(x)\psicha_{r',\kk'}(x).
\label{appE:Jexpr}
\ee

\bigskip
The Hamiltonian is of the usual form
\be
\hat H&=&\Hph+\Hel+\HI
\ee
with
\be
\Hph&=&\hbar c|\kkph|\left(\ah^\dagger\ah+\av^\dagger\av\right),\\
\Hel&=&\hbar\omega(\kk)\sum_{s=1}^4\hat\phi^{(-)}_{s}\hat\phi^{(+)}_{s},\\
\HI&=&\int_{\Ro^3}\upd \xx\,\hat j^\mu(\xx,0) \hat A_\mu(\xx,0).
\ee
The interaction Hamiltonian can be written out as
\be
\HI\psi_{\kkph,\kk}
&=&
\int\upd \xx\,
\frac{qc}{(2\pi)^3}\int\upd \kk'\,
\hat A_{\mu,\kkph}(x)\hat J^\mu_{\kk,\kk'}(x)\psi_{\kkph,\kk'}\bigg|_{x^0=0}.
\label{hamil:HI}
\ee
Its expectation satisfies
\be
\langle\psi_{\kkph,\kk}|\HI\psi_{\kkph',\kk'}\rangle
&=&
\frac{qc}{(2\pi)^3}\int\upd \xx\,\int\upd k'\,
\langle \psi_{\kkph,\kk}|\hat A_{\mu,\kkph}(x)\hat J^\mu_{\kk,\kk'}(x)\psi_{\kkph,\kk'}\rangle\cr
&=&
\overline{\langle\psi_{\kkph',\kk'}|\HI\psi_{\kkph,\kk}\rangle}.
\ee
To prove this use that the operators $\hat A_{\mu,\kkph}(x)$ and $\hat J^\mu_{\kk,\kk'}(x)$
commute with each other.
The total charge
\be
\hat Q&=&\frac 1c\int\upd \xx \,\hat j^0(x)
\label{electron:totcharge}
\ee
is conserved because (see (II))
\be
\left[\hat J^\mu_{\kk,\kk'}(x),\hat Q\right]_-&=&0.
\ee

\subsection{The emergent picture}

In \cite{CM79} a unitary transformation $\hat V$ is defined by a generator $\hat T(\xx)$
through
\be
\hat V&=&\exp\left(i\int\upd^3 \xx\,\hat T(\xx)\right).
\ee
The generator is of the form
\be
\hat T(\xx)&=&\frac{q}{4\pi}\int\upd^3 \yy\,\hat \Abf(x)\cdot\frac{\xx-\yy}{|\xx-\yy|^3}\hat j_0(\yy).
\ee
Here, $q$ is the elementary unit of charge. Bold characters are used to indicate three-vectors.
The result of \cite{CM79}, in the context of standard QED, is that 
\be
\hat V\nabla\cdot\hat\Ebf\hat V^{-1}&=&\nabla\cdot\hat\Ebf-q\hat j_0,
\ee
where $\hat\Ebf(x)$ are the electric field operators. If they satisfy Gauss's law
in the presence of a charge distribution $\hat j_0(\xx)$
then $\hat V\nabla\cdot\hat\Ebf\hat V^{-1}$ satisfies Gauss's law in absence of charges.

\bigskip

In the present work the trick of \cite{CM79} is directly applied to define new electric
field operators. They are defined by
\be
& &\hat E''_\alpha(x)=\hat E'_\alpha(x)
+\frac{\mu_0 c}{4\pi}\frac{\partial\,}{\partial x^\alpha}\int\upd\yy\,
\frac{1}{|\xx-\yy|}\,\hat U(-x^0)\hat j^0(\yy,0)\hat U(x^0).\cr
& &
\label{emerg:def2}
\ee
Here, 
\be
\hat U(x^0)=\exp(-ix^0\hat H/\hbar c)
\label{emerg:unitary}
\ee
is the time evolution of the interacting system.
The new operators are marked with a double prime to distinguish them from the
operators of the non-interacting system (without a prime)
and those of the interacting system (with a single prime).
One verifies immediately that Gauss' law is satisfied
\be
\sum_\alpha\frac{\partial\,}{\partial x^\alpha}\hat E''_\alpha(x)
&=&
-\mu_0 c\,\hat j^{0\prime}(x).
\ee
This follows because the Coulomb potential is minus the Green's function of the Laplacian.

\bigskip

The second term in the r.h.s.~of (\ref {emerg:def2}) is the Coulomb contribution to the electric field.
The curl of this term vanishes. Hence it is obvious to take
\be
\hat B''_\alpha(x)&\equiv& \hat B'_\alpha(x).
\ee
This implies the second of the four equations of Maxwell,
stating that the divergence of $\hat B''_\alpha(x)$ vanishes.
Also the fourth equation, absence of magnetic charges,
follows immediately because $\hat E''(x)$ and $\hat E'(x)$
have the same curl. Remains to write Faraday's law as
\be
(\nabla\times\hat B''(x))_\alpha-\frac 1{c}\frac{\partial\,}{\partial x^0}\hat E''_\alpha(x)
&=&
-\mu_0\,\hat j''_\alpha(x)
\ee
with
\be
\hat j''_\alpha(x)
&=&
-\frac 1{\mu_0 c}\frac{\partial\,}{\partial x^0}\left(\hat E''_\alpha(x)-\hat E'_\alpha(x)\right).
\label{emerg:current}
\ee
Finally, take $\hat j''_0(x)=\hat j'_0(x)$.
A short calculation shows that the newly defined current operators $\hat j''_\mu(x)$
satisfy the continuity equation.
Note that the operators $\hat j''_\alpha(x)$, $\alpha=1,2,3,$
are fully determined by the charge field $\hat j'_0(x)$.

\subsection{Discussion}

The operators $\hat E''_\alpha(x)$, $\hat B''(x))_\alpha$, $\hat j''_\mu(x)$, form what I call
the {\em emergent picture} of QED. Their time evolution is the same as that of the operators
in the original Heisenberg picture and is determined by the unitary operators (\ref {emerg:unitary}).
Because of $\hat j''_0(x)=\hat j'_0(x)$ also the time-dependent charge field is the same
in the two pictures.

\bigskip
Is there a physical interpretation underlying the mathematical equivalence of the two pictures?
Here the analogy with the polaron comes into sight. A first step underpinning this analogy
is the proof of the existence of bound states, which follows in the next section.

%%%%%%%%%%%%%%%%%%%%%%%%%%%%%%%%%%%%
%%%%%%%%%%%%%%%%%%%%%%%%%%%%%%%%%%%%
%%%%%%%%%%%%%%%%%%%%%%%%%%%%%%%%%%%%
%%%%%%%%%%%%%%%%%%%%%%%%%%%%%%%%%%%%
%%%%%%%%%%%%%%%%%%%%%%%%%%%%%%%%%%%%
\section{Bound states}
\label{sect:bound}

\subsection{Trial wave functions}
\label{subsect:trial}

Consider a wave function of the form
\be
\psi_{\kkph,\kk}
&=&
\sum_{m,n=0}^\infty\tau_{m,n}(\kkph,\kk)\frac{\sqrt{\rho(\kkph)\rhoel(\kk)}}{\sqrt{m!n!\,\Zel(\kk)}}
\,|m,n\rangle\times|\{1\}\rangle\cr
& &
+\sqrt{1-\rho(\kkph)\rhoel(\kk)}\,|0,0\rangle\times|\emptyset\rangle\cr
& &
\ee
with $\tau_{m,n}(\kkph,\kk)$ either 1 or 0,
and with $\rho(\kkph)\rhoel(\kk)\le 1$  for all $\kkph$, $\kk$.
Let us assume that
\be
\sum_{m,n=0}^\infty\frac{1}{m!n!}\tau_{m,n}(\kkph,\kk)&=&\Zel(\kk)
\ee
holds independently of the value of $\kkph$ and that
\be
\lu^3\int\upd\kkph\,\rho(\kkph)&=&1.
\label{trial:normalized}
\ee
The wave function is properly normalized.
It describes a superposition of vacuum with a spin-up electron field entangled with
a photon field.

\bigskip

The different contributions to the total energy are as follows.
The kinetic energy of the photon field is
\be
\langle\Hph\rangle
&=&
\lu^6\int\upd\kkph\,\int\upd\kk\,\hbar c|\kkph|\sum_{m,n=0}^\infty\frac{m+n}{m!n!}
\tau_{m,n}(\kkph,\kk)\rho(\kkph)\frac{\rhoel(\kk)}{\Zel(\kk)}\cr
&=&
\lu^3\int\upd\kkph\,\hbar c|\kkph|\rho(\kkph)\Zph(\kkph)
\ee
with
\be
\Zph(\kkph)&=&\lu^3\int\upd\kk\,
\frac{\rhoel(\kk)}{\Zel(\kk)}\sum_{m,n=0}^\infty\frac{m+n}{m!n!}\tau_{m,n}(\kkph,\kk).
\ee
The kinetic energy of the electron field is
\be
\langle\Hel\rangle
&=&
\lu^3\int\upd\kk\,\hbar\omega(\kk)\rhoel(\kk).
\ee
The interaction energy equals
\be
\langle\HI\rangle
&=&
\lu^6\int\upd\kkph\int\upd\kk\,\langle\psi|\HI\psi\rangle\cr
&=&
\frac{qc}{(2\pi)^3}\lu^6\int\upd\kkph\,\rho(\kkph)\int\upd\kk\,\int\upd \kk'\,
\int\upd \xx\,
\frac{\sqrt{\rhoel(\kk)\rhoel(\kk')}}{\sqrt{\Zel(\kk)\Zel(\kk')}}\cr
& &\times
\sum_{m,m',n,n'=0}^\infty\frac{\tau_{m,n}(\kkph,\kk)\tau_{m',n'}(\kkph,\kk')}{\sqrt{m!m'!n!n'!}}\cr
& &\quad\times
\sum_\alpha
\langle m,n|\hat A_{\alpha,\kkph}(\xx,0)|m',n'\rangle\,
\langle\{1\}|\hat J^\alpha_{\kk,\kk'}(\xx,0)|\{1\}\rangle
\cr
&=&
-\lu^3\int\upd\kkph\,\rho(\kkph)w(\kkph)
\ee
with
\be
w(\kkph)
&=&
-\frac{qc}{(2\pi)^3}\lu^3\int\upd\kk\,\int\upd \kk'\,
\frac{\sqrt{\rhoel(\kk)\rhoel(\kk')}}{\sqrt{\Zel(\kk)\Zel(\kk')}}\cr
& &\times
\sum_{m,m',n,n'=0}^\infty\frac{\tau_{m,n}(\kkph,\kk)\tau_{m',n'}(\kkph,\kk')}{\sqrt{m!m'!n!n'!}}\cr
& &\quad\times
\int\upd \xx\,
\sum_\alpha
\langle m,n|\hat A_{\alpha,\kkph}(\xx,0)|m',n'\rangle\,
\langle\{1\}|\hat J^\alpha_{\kk,\kk'}(\xx,0)|\{1\}\rangle.
\label{bound:wdef}
\ee

\bigskip

It is shown in the Appendix \ref {app:intenerg} that
the function $w(\kkph)$ can be written as 
\be
w(\kkph)
&=&
-\lu^3\int\upd\kk\,U^{(H)}(\kkph,\kk)
\sum_{m,n=0}^\infty\frac{1}{m!n!}\bigg\{
\frac {\tau_{m,n}(\kkph,\kk)\tau_{m+1,n}(\kkph,\kk+\kkph)}
{\sqrt{\Zel(\kk)\Zel(\kk+\kkph)}}\cr
& &\qquad\qquad
-\frac{\tau_{m,n}(\kkph,-\kk)\tau_{m+1,n}(\kkph,-\kk-\kkph)}
{\sqrt{\Zel(-\kk)\Zel(-\kk-\kkph)}}
\bigg\}\cr
& &
-\lu^3\int\upd\kk\,U^{(V)}(\kkph,\kk)
\sum_{m,n=0}^\infty\frac{1}{m!n!}\bigg\{
\frac {\tau_{m,n}(\kkph,\kk)\tau_{m,n+1}(\kkph,\kk+\kkph)}
{\sqrt{\Zel(\kk)\Zel(\kk+\kkph)}}\cr
& &\qquad\qquad
+\frac{\tau_{m,n}(\kkph,-\kk)\tau_{m,n+1}(\kkph,-\kk-\kkph)}
{\sqrt{\Zel(-\kk)\Zel(-\kk-\kkph)}}
\bigg\}\cr
& &
\label{bound:wexpr}
\ee
with functions $U^{(H/V)}(\kkph,\kk)$ given by
\be
U^{(H/V)}(\kkph,\kk)
&=&
\frac 12qc\lambda\sqrt{\rhoel(\kk)\rhoel(\kk+\kkph)}\sum_\alpha \varepsilon^{(H/V)}_\alpha(\kkph)
\Re \langle u^{(1)}(\kk+\kkph)|\gamma^0\gamma^\alpha u^{(1)}(\kk)\rangle.\cr
& &
\label{bound:UHVdef}
\ee
The symmetries of these functions are important for what follows.
From $\varepsilon^{(H)}_\alpha(-\kkph)=\varepsilon^{(H)}_\alpha(\kkph)$
and $\varepsilon^{(V)}_\alpha(-\kkph)=-\varepsilon^{(V)}_\alpha(\kkph)$
and 
\be
\langle u^{(1)}(-\kk')|\gamma^0\gamma^\alpha u^{(1)}(-\kk)\rangle
&=&
-\langle u^{(1)}(\kk')|\gamma^0\gamma^\alpha u^{(1)}(\kk)\rangle
\ee
follows that $U^{(H)}(-\kkph,-\kk)=-U^{(H)}(\kkph,\kk)$ and $U^{(V)}(-\kkph,-\kk)=U^{(V)}(\kkph,\kk)$.
In addition is
$U^{(H)}(-\kkph,\kk)=U^{(H)}(\kkph,\kk-\kkph)$
and $U^{(V)}(-\kkph,\kk)=-U^{(V)}(\kkph,\kk-\kkph)$.
It is shown in the Appendix \ref {app:estim} that
\be
U^{(H/V)}(\kkph,\kk)\le 0\quad\mbox{ if and only if }\quad 
\sum_\alpha  \varepsilon^{(H/V)}_\alpha(\kkph)k_\alpha\ge 0.
\label{estim:UHVsign}
\ee
This implies in particular that $U^{(H/V)}(\kkph,\kk)$ and $U^{(H/V)}(\kkph,\kk\pm\kkph)$
have the same sign.

%%%%%%%%%%%%%%%%%%%%%%%%%%%%%%%%%%%%
\subsection{A specific choice}

Let us now make a simple choice for the coefficients $\tau_{m,n}$
\be
\tau_{1,0}(\kkph,\kk)&=&\Theta(\kk\cdot\varepsilon^{(H)}(\kkph)),\\
\tau_{0,0}(\kkph,\kk)&=&1-\tau_{1,0}(\kkph,\kk),\\
\tau_{m,n}(\kkph,\kk)&=&0\quad\mbox{ otherwise}.
\ee
Then $\Zel(\kk)=1$ for all $\kk$.
Note that $\tau_{1,0}(\kkph,\kk\pm\kkph)=\tau_{1,0}(\kkph,\kk)$.
The expression (\ref {bound:wexpr}) for $w(\kkph)$ becomes
\be
w(\kkph)
&=&
-\lu^3\int\upd\kk\,
U^{(H)}(\kkph,\kk)\{1-\tau_{1,0}(\kkph,\kk)\}
\bigg\{
\tau_{1,0}(\kkph,\kk)
-\tau_{1,0}(\kkph,-\kk)
\bigg\}\cr
&=&
\lu^3\int\upd\kk\,
U^{(H)}(\kkph,\kk)\tau_{1,0}(\kkph,-\kk)\cr
&=&
\lu^3\int\upd\kk\,
U^{(H)}(\kkph,-\kk)\tau_{1,0}(\kkph,\kk).
\label{choice:wexpr}
\ee

\bigskip

Assume that $\rhoel(\kk)$ has a Gaussian shape
\be
\rhoel(\kk)
&=&\rhoel(0)\exp\left(-\frac{1}{2\sigma^2}|\kk|^2\right).
\ee
Then it is easy to evaluate the long wavelength limit of the function $U^{(H)}(\kkph,\kk)$
(see the Appendix \ref{app:specific}). The result is
\be
U^{(H/V)}(\kkph,\kk)
&=&
-\frac 12qc\lambda\rhoel(\kk)\left[1-\frac{\kk\cdot\kkph}{2\sigma^2}+\mbox{O}(|\kkph|^2)\right]\cr
& &\times
\frac{2}{\omega(\kk)}\left[1+\frac{1}{2}\frac{\kk\cdot\kkph}{\mstar^2+|\kk|^2}+\mbox{O}(|\kkph|^2)\right]
\,\kk\cdot  \varepsilon^{(H/V)}(\kkph).
\label{specific:expfun}
\ee
There follows
\be
w(\kkph)-\hbar c|\kkph|\Zph(\kkph)
&=&
\lu^3\int\upd\kk\,\Theta(\kk\cdot\varepsilon^{(H)}(\kkph))
\bigg[
\frac 12qc\lambda\rhoel(\kk)\left[1+\frac{\kk\cdot\kkph}{2\sigma^2}+\mbox{O}(|\kkph|^2)\right]\cr
& &\times
\frac{2}{\omega(\kk)}\left[1-\frac{1}{2}\frac{\kk\cdot\kkph}{\mstar^2+|\kk|^2}+\mbox{O}(|\kkph|^2)\right]
\,\kk\cdot  \varepsilon^{(H)}(\kkph)\cr
& &
-\hbar c|\kkph|\rhoel(\kk)
\bigg].
\ee
Because of mirror symmetry in the direction of $\kkph$
the terms in $\kk\cdot\kkph$ drop. The result is
\be
w(\kkph)-\hbar c|\kkph|\Zph(\kkph)
&=&
\lu^3\int\upd\kk\,\Theta(\kk\cdot\varepsilon^{(H)}(\kkph))\,\rhoel(\kk)\cr
& &\times
\left[
qc\lambda
\frac{1}{\omega(\kk)}\,[\kk\cdot\varepsilon^{(H)}(\kkph)]
-\hbar c|\kkph|
+\mbox{O}(|\kkph|^2\right].
\ee
This suggests the following

\begin{lemma}
\label{lemma:binding}
There exists $\epsilon>0$, independent of $\rhoel(0)$, such that
 \be
\mbox{ if }
w(\kkph)-\hbar c|\kkph|\Zph(\kkph)>0
\quad\mbox{ whenever }\quad
|\kkph|\sigma\exp(|\kkph|^2/4\sigma^2)<\epsilon
\label{long:lemmacon}
\ee
then $ w(\kkph)-\hbar c|\kkph|\Zph(\kkph)>0$.

\end{lemma}

\noindent
A proof of the Lemma is found in the Appendix \ref {app:prooflemma}.

\bigskip

Use the Lemma to estimate
\be
\langle\Hph\rangle+\langle\HI\rangle
&=&
-\lu^3\int\upd\kkph\,\rho(\kkph)
\left[w(\kkph)-\hbar c|\kkph|Z(\kkph)\right].
\label{long:temp}
\ee
Take $\rho(\kkph)=0$ when $|\kkph|>\delta$
where $\delta$ is the solution of 
\be
\delta\sigma\exp(\delta^2/4\sigma^2)=\epsilon.
\ee
The constraints (\ref {trial:normalized}) and 
\be
\rho(\kkph)\rhoel(\kk)\le 1\quad\mbox{ for all }\kkph, \kk,
\ee
can be satisfied by taking
\be
\rho(\kkph)&=&\frac{3}{4\pi\delta^2\lu^3},\qquad |\kkph|<\delta,
\ee
and requiring that
\be
0<\rhoel(0)\le\frac{4\pi}{3}\delta^3\lu^3.
\ee
Then the conditions of the Lemma are satisfied.
One concludes that the integrand in the r.h.s.~of (\ref {long:temp})
cannot be negative, and hence that the l.h.s.~is strictly negative.
This proves the following result

\begin{theorem}
Let  \quad $\rhoel(\kk)=\rhoel(0)\exp(-|\kk|^2/2\sigma^2)$\quad  with $\rhoel(0)>0$ but small.
There exist wave functions $\psi$ for which 
\be
\langle\Hph\rangle+\langle\HI\rangle
&<&0\qquad\mbox{ and}\\
\int\upd\kkph\, \langle\psi|\Hel\psi\rangle_{\kkph,\kk}&=&\rhoel(\kk)\,\hbar\omega(\kk).
\ee

\end{theorem}

%%%%%%%%%%%%%%%%%%%%%%%%%%%%%%%%%%%%
\subsection{Discussion}

The Theorem shows that the energy of an electron field can be lowered by binding with an electromagnetic field
which is transversely polarized. Because of the conservation law the total energy remains at all times 
lower than the initial kinetic energy $\langle\Hel\rangle$ of the electron field.
However, one cannot exclude that part of the kinetic energy of the electron field is converted into
asymptotically free photons, for instance by means of Bremsstrahlung.
The open question is therefore whether the dressed electron field looses all of its dressing
as time progresses, or keeps part of it.
To answer this question the time evolution of the interacting system has to be studied.
This is out of the scope of the present paper.

\bigskip

The proof of the Theorem uses a particular trial wave function.
Of course, other choices are possible. The more general setup of Section \ref {subsect:trial}
allows to construct further examples.
However, a more systematic investigation is needed. In particular one wants to know
the optimal choice of wave function and whether the total energy remains finite.

\bigskip

Note that only low photon numbers are relevant because the cost of
creating an electromagnetic field increases linearly with the number of photons while the 
electromagnetic field strength is proportional to the square root.
Note also that the choice of trial function depends on the spin of the electron field.

%%%%%%%%%%%%%%%%%%%%%%%%%%%%%%%%%%%%
%%%%%%%%%%%%%%%%%%%%%%%%%%%%%%%%%%%%
%%%%%%%%%%%%%%%%%%%%%%%%%%%%%%%%%%%%
%%%%%%%%%%%%%%%%%%%%%%%%%%%%%%%%%%%%
%%%%%%%%%%%%%%%%%%%%%%%%%%%%%%%%%%%%
\section{Final Remarks}

In (I) electromagnetic potential operators $\hat A_\alpha(x)$ are
introduced. They describe free photon fields in the temporal gauge.
In (II) the Dirac current operators $\hat j^\mu(x)$ are defined
for a non-interacting electron-positron field. Together, these
operators $\hat A_\alpha(x)$ and $\hat j^\mu(x)$ determine the
interaction Hamiltonian $\HI$. It is given by (\ref {hamil:HI}).
The operators of the interacting system are marked with a prime.
The electric field operators $\hat E'_\alpha(x)$ still satisfy Gauss' law
in absence of a charge field. However, the simple transformation (\ref {emerg:def2})
introduces new field operators which are such that Gauss' law now reproduces
the given charge field $\hat j'_0(x)$. This leads to the conclusion
that two equivalent descriptions of the same time-dependent charge field exist,
one in which Coulomb forces are absent, one in which they emerge as an
effective description. This observation justifies the study of a version of
QED in which no longitudinal or scalar photons exist.

\bigskip

Leaving out the longitudinal and scalar photons eliminates an important source of mathematical
inconsistencies. The resulting theory can be formulated in a mathematically rigorous fashion.
However, the use of excessive mathematical machinery has been avoided here to
keep the paper readable for a larger audience.

\bigskip
An analogy with the polaron problem is made to explain the emergence of Coulomb forces
as an effective description of the time dependence of the charge
field. As a first step in revealing the nature of the interactions
the existence of bound states is established.
An analysis of the time evolution is still missing.

\bigskip
Many open questions remain.
For instance, polarons do attract each other, while regions where the charge of the electron field has
equal sign should repel. In addition the forces should decay inversely proportional to the
square of the distance. Hopeful in this respect is that the binding of the photon field is
maximal in the long wavelength limit. This supports at least qualitatively that
the forces are long-ranged.

\bigskip

The next paper in this series of papers treats scattering theory in the 
context of reducible QED. In particular, it focuses on 
the consequences of omitting longitudinal and scalar photons.

%%%%%%%%%%%%%%%%%%%%%%%%%%%%%%%%%%%%
%%%%%%%%%%%%%%%%%%%%%%%%%%%%%%%%%%%%
%%%%%%%%%%%%%%%%%%%%%%%%%%%%%%%%%%%%
%%%%%%%%%%%%%%%%%%%%%%%%%%%%%%%%%%%%
%%%%%%%%%%%%%%%%%%%%%%%%%%%%%%%%%%%%
\appendix
\section*{Appendices}

\section{Calculation of the interaction energy}
\label{app:intenerg}

Let us evaluate the function $w(\kkph)$ as defined by (\ref {bound:wdef}).
Use
\be
\langle m,n|\hat A_{\alpha,\kkph}(\xx,0)|m',n'\rangle
&=&
\frac 1{2}\lambda\varepsilon^{(H)}_\alpha(\kkph)
\langle m,n|\left[e^{i\kkph\cdot\xx}\ah+e^{-i\kkph\cdot\xx}\ah^\dagger\right]|m',n'\rangle\cr
& &+\frac 1{2}\lambda\varepsilon^{(V)}_\alpha(\kkph)
\langle m,n|\left[e^{i\kkph\cdot\xx}\av+e^{-i\kkph\cdot\xx}\av^\dagger\right]|m',n'\rangle\cr
&=&
\frac 1{2}\lambda\varepsilon^{(H)}_\alpha(\kkph)
\left[e^{i\kkph\cdot\xx}\sqrt{m+1}\delta_{m+1,m'}
+e^{-i\kkph\cdot\xx}\sqrt{m'+1}\delta_{m'+1,m}\right]\delta_{n,n'}\cr
& &
+\frac 1{2}\lambda\varepsilon^{(V)}_\alpha(\kkph)\delta_{m,m'}
\left[e^{i\kkph\cdot\xx}\sqrt{n+1}\delta_{n+1,n'}+e^{-i\kkph\cdot\xx}\sqrt{n'+1}\delta_{n,n'+1}\right].\cr
& &
\ee
This gives
\be
w(\kkph)
&=&
-\frac{qc\lambda}{2(2\pi)^3}\lu^3\int\upd\kk\,\int\upd \kk'\,
\frac{\sqrt{\rhoel(\kk)\rhoel(\kk')}}{\sqrt{\Zel(\kk)\Zel(\kk')}}\cr
& &\times
\bigg\{
\sum_{m,m',n=0}^\infty\frac{\tau_{m,n}(\kkph,\kk)\tau_{m',n}(\kkph,\kk')}{\sqrt{m!m'!}\,n!}\cr
& &\qquad\times
\int\upd \xx\,
\sum_\alpha \varepsilon^{(H)}_\alpha(\kkph)
\left[e^{i\kkph\cdot\xx}\sqrt{m+1}\delta_{m+1,m'}
+e^{-i\kkph\cdot\xx}\sqrt{m'+1}\delta_{m'+1,m}\right]\cr
& &\quad
+
\sum_{m,n,n'=0}^\infty\frac{\tau_{m,n}(\kkph,\kk)\tau_{m,n'}(\kkph,\kk')}{m!\sqrt{n!n'!}}\cr
& &\qquad\times
\int\upd \xx\,
\sum_\alpha\varepsilon^{(V)}_\alpha(\kkph)
\left[e^{i\kkph\cdot\xx}\sqrt{n+1}\delta_{n+1,n'}+e^{-i\kkph\cdot\xx}\sqrt{n'+1}\delta_{n,n'+1}\right]\cr
& &
\bigg\}\langle\{1\}|\hat J^\alpha_{\kk,\kk'}(\xx,0)|\{1\}\rangle\cr
&=&
-\frac{qc\lambda}{2(2\pi)^3}\lu^3\int\upd\kk\,\int\upd \kk'\,\int\upd \xx\,
\frac{\sqrt{\rhoel(\kk)\rhoel(\kk')}}{\sqrt{\Zel(\kk)\Zel(\kk')}}\cr
& &\times
\bigg\{
\sum_{m,n=0}^\infty\frac{\tau_{m,n}(\kkph,\kk)\tau_{m+1,n}(\kkph,\kk')}{m!n!}
\sum_\alpha \varepsilon^{(H)}_\alpha(\kkph)
e^{i\kkph\cdot\xx}\cr
& &\quad
+\sum_{m',n=0}^\infty\frac{\tau_{m'+1,n}(\kkph,\kk)\tau_{m',n}(\kkph,\kk')}{m'!n!}
\sum_\alpha \varepsilon^{(H)}_\alpha(\kkph)
e^{-i\kkph\cdot\xx}\cr
& &\quad
+
\sum_{m,n=0}^\infty\frac{\tau_{m,n}(\kkph,\kk)\tau_{m,n+1}(\kkph,\kk')}{m!n!}
\sum_\alpha\varepsilon^{(V)}_\alpha(\kkph)
e^{i\kkph\cdot\xx}\cr
& &\quad
+
\sum_{m,n'=0}^\infty\frac{\tau_{m,n'+1}(\kkph,\kk)\tau_{m,n'}(\kkph,\kk')}{m!n'!}
\sum_\alpha\varepsilon^{(V)}_\alpha(\kkph)
e^{-i\kkph\cdot\xx}\cr
& &
\bigg\}\langle\{1\}|\hat J^\alpha_{\kk,\kk'}(\xx,0)|\{1\}\rangle.\cr
&&
\ee
Use that $\hat J^\alpha_{\kk,\kk'}(x)=\hat J^\alpha_{\kk',\kk}(x)$ to obtain
\be
w(\kkph)
&=&
-\frac{qc\lambda}{(2\pi)^3}\lu^3\int\upd\kk\,\int\upd \kk'\,
\frac{\sqrt{\rhoel(\kk)\rhoel(\kk')}}{\sqrt{\Zel(\kk)\Zel(\kk')}}
\int\upd \xx\,\cos(\kkph\cdot\xx)
\sum_{m,n=0}^\infty\frac{1}{m!n!}\cr
& &\times
\bigg\{
\tau_{m,n}(\kkph,\kk)\tau_{m+1,n}(\kkph,\kk')
\sum_\alpha \varepsilon^{(H)}_\alpha(\kkph)\langle\{1\}|\hat J^\alpha_{\kk,\kk'}(\xx,0)|\{1\}\rangle
\cr
& &\quad
+\tau_{m,n}(\kkph,\kk)\tau_{m,n+1}(\kkph,\kk')
\sum_\alpha\varepsilon^{(V)}_\alpha(\kkph)\langle\{1\}|\hat J^\alpha_{\kk,\kk'}(\xx,0)|\{1\}\rangle
\bigg\}.\cr
&&
\ee
Now use that
\be
& &
\frac{1}{(2\pi)^3}
\int\upd \xx\,e^{\pm i\kkph\cdot\xx}\,\langle\{1\}|\hat J^\alpha_{\kk,\kk'}(\xx,0)|\{1\}\rangle\cr
&=&
\frac{1}{(2\pi)^3}\int\upd \xx\,e^{\pm i\kkph\cdot\xx}
\frac 12\sum_{s,t=1,2}
\langle u^{(s)}(\kk)|\gamma^0\gamma^\alpha u^{(t)}(\kk')\rangle\,
\langle\{1\}|\hat\phi_{s,\kk}^{(-)}(\xx,0)\hat\phi_{t,\kk'}^{(+)}(\xx,0)|\{1\}\rangle
\cr
& &
+\frac{1}{(2\pi)^3}\int\upd \xx\,e^{\pm i\kkph\cdot\xx}
\frac 12\sum_{s,t=1,2}
\langle u^{(s)}(\kk')|\gamma^0\gamma^\alpha u^{(t)}(\kk)\rangle\,
\langle\{1\}|\hat\phi_{s,\kk'}^{(-)}(\xx,0)\hat\phi_{t,\kk}^{(+)}(\xx,0)|\{1\}\rangle\cr
&=&
\frac{1}{(2\pi)^3}\int\upd \xx\,e^{\pm i\kkph\cdot\xx}e^{-i(\kk-\kk')\cdot\xx}
\frac 12
\langle u^{(1)}(\kk)|\gamma^0\gamma^\alpha u^{(1)}(\kk')\rangle\cr
& &
+\frac{1}{(2\pi)^3}\int\upd \xx\,e^{\pm i\kkph\cdot\xx}e^{i(\kk-\kk')\cdot\xx}
\frac 12
\langle u^{(1)}(\kk')|\gamma^0\gamma^\alpha u^{(1)}(\kk)\rangle\cr
&=&
\frac 12\delta(\kk-\kk'\mp\kkph)
\langle u^{(1)}(\kk)|\gamma^0\gamma^\alpha u^{(1)}(\kk')\rangle\cr
& &
+\frac 12\delta(\kk-\kk'\pm\kkph)
\langle u^{(1)}(\kk')|\gamma^0\gamma^\alpha u^{(1)}(\kk)\rangle.
\ee
This gives
\be
w(\kkph)
&=&
-qc\lambda\lu^3\int\upd\kk\,\int\upd \kk'\,
\frac{\sqrt{\rhoel(\kk)\rhoel(\kk')}}{\sqrt{\Zel(\kk)\Zel(\kk')}}
\frac 12\left[\delta(\kk-\kk'+\kkph)+\delta(\kk-\kk'-\kkph)\right]
\cr
& &\times
\sum_{m,n=0}^\infty\frac{1}{m!n!}\bigg\{
\tau_{m,n}(\kkph,\kk)\tau_{m+1,n}(\kkph,\kk')
\sum_\alpha \varepsilon^{(H)}_\alpha(\kkph)\Re \langle u^{(1)}(\kk')|\gamma^0\gamma^\alpha u^{(1)}(\kk)\rangle
\cr
& &\qquad\qquad
+\tau_{m,n}(\kkph,\kk)\tau_{m,n+1}(\kkph,\kk')
\sum_\alpha\varepsilon^{(V)}_\alpha(\kkph)
\Re \langle u^{(1)}(\kk')|\gamma^0\gamma^\alpha u^{(1)}(\kk)\rangle
\bigg\}\cr
&=&
-\lu^3\int\upd\kk\,
\sum_{m,n=0}^\infty\frac{1}{m!n!}\bigg\{
\tau_{m,n}(\kkph,\kk)\tau_{m+1,n}(\kkph,\kk+\kkph)
\frac {U^{(H)}(\kkph,\kk)}{\sqrt{\Zel(\kk)\Zel(\kk+\kkph)}}\cr
& &\qquad\qquad
+\tau_{m,n}(\kkph,\kk)\tau_{m+1,n}(\kkph,\kk-\kkph)
\frac{U^{(H)}(-\kkph,\kk)}{\sqrt{\Zel(\kk)\Zel(\kk-\kkph)}}\cr
& &\qquad\qquad
+\tau_{m,n}(\kkph,\kk)\tau_{m,n+1}(\kkph,\kk+\kkph)
\frac {U^{(V)}(\kkph,\kk)}{\sqrt{\Zel(\kk)\Zel(\kk+\kkph)}}\cr
& &\qquad\qquad
+\tau_{m,n}(\kkph,\kk)\tau_{m,n+1}(\kkph,\kk-\kkph)
\frac{U^{(V)}(-\kkph,\kk)}{\sqrt{\Zel(\kk)\Zel(\kk-\kkph)}}
\bigg\}.\cr
& &
\ee
One can write this result as (\ref{bound:wexpr}).

\section{Estimates}
\label{app:estim}

One has
\be
\langle u^{(1)}(\kk+\kkph)|\gamma^0\gamma^\alpha u^{(1)}(\kk)\rangle
&=&
-c\frac{k_\alpha[\omega(\kk+\kkph)+c\mstar]+(k_\alpha+\kph_\alpha)[\omega(\kk)+c\mstar]}
{\sqrt{\omega(\kk+\kkph)[\omega(\kk+\kkph)+c\mstar]}\,{\sqrt{\omega(\kk)[\omega(\kk)+c\mstar]}}}.
\ee
This implies
\be
\sum_\alpha \varepsilon^{(H/V)}_\alpha(\kkph)
\Re \langle u^{(1)}(\kk+\kkph)|\gamma^0\gamma^\alpha u^{(1)}(\kk)\rangle
&=&
-\sum_\alpha  \varepsilon^{(H/V)}_\alpha(\kkph)\left[Ak_\alpha+B\kph_\alpha\right],
\ee
\be
\mbox{with }\quad
A
&=&
c\frac{\omega(\kk)+\omega(\kk+\kkph)+2c\mstar}
{\sqrt{\omega(\kk+\kkph)[\omega(\kk+\kkph)+c\mstar]}\,{\sqrt{\omega(\kk)[\omega(\kk)+c\mstar]}}}
\label{estim:Adef}
\\
\mbox{and }\quad
B
&=&
c\frac{\omega(\kk)+c\mstar}
{\sqrt{\omega(\kk+\kkph)[\omega(\kk+\kkph)+c\mstar]}\,{\sqrt{\omega(\kk)[\omega(\kk)+c\mstar]}}}.
\ee
Because $\sum_\alpha  \varepsilon^{(H/V)}_\alpha(\kkph)\kph_\alpha=0$ there follows
\be
\sum_\alpha \varepsilon^{(H/V)}_\alpha(\kkph)
\Re \langle u^{(1)}(\kk+\kkph)|\gamma^0\gamma^\alpha u^{(1)}(\kk)\rangle
&=&
-A\sum_\alpha  \varepsilon^{(H/V)}_\alpha(\kkph)k_\alpha. 
\label{est:result}
\ee
One concludes that (\ref {estim:UHVsign}) holds.

\section{Long wavelength limit}
\label {app:specific}

Here, the long wavelength expansion of the function $U^{(H)}(\kkph,\kk)$
is calculated.

\bigskip

From (\ref {est:result}) and
\be
A&=&
\frac{2}{\omega(\kk)}\left[1+\frac{1}{2}\frac{\kk\cdot\kkph}{\mstar^2+|\kk|^2}+\mbox{O}(|\kkph|^2)\right]
\ee
follows
\be
& &
\sum_\alpha \varepsilon^{(H/V)}_\alpha(\kkph)
\Re \langle u^{(1)}(\kk+\kkph)|\gamma^0\gamma^\alpha u^{(1)}(\kk)\rangle\cr
&=&
-\frac{2}{\omega(\kk)}\left[1+\frac{1}{2}\frac{\kk\cdot\kkph}{\mstar^2+|\kk|^2}+\mbox{O}(|\kkph|^2)\right]
% \cr
% & &\times
\sum_\alpha  \varepsilon^{(H/V)}_\alpha(\kkph)k_\alpha.
\label{longapp:temp}
\ee

\bigskip

From the assumption that $\rhoel(\kk)$ has a Gaussian shape follows
\be
\rhoel(\kk+\kkph)&=&\rhoel(\kk)\left[1-\frac{\kk\cdot\kkph}{\sigma^2}+\mbox{O}(|\kkph|^2)\right].
\ee
Combining this expression with (\ref {longapp:temp}) yields (\ref {specific:expfun}).

\section{Proof of the Lemma}
\label{app:prooflemma}

Here a proof of Lemma \ref {lemma:binding} follows.

Assume $|\kk|<\mstar$ and $|\kkph|<\mstar$ and $\kk\cdot \varepsilon^{(H)}(\kkph)>0$.
The quantity $A$ defined by (\ref {estim:Adef}) satisfies
\be
A>\frac{4}{\sqrt{\sqrt 5(1+\sqrt 5)\sqrt 2(1+\sqrt 2}}\,\frac{1}{\mstar}>
\frac{1}{7\mstar}.
\ee
From (\ref {est:result}) and (\ref {bound:UHVdef}) then follows
\be
U^{(H)}(\kkph,-\kk)
&>&
\frac 1{14\mstar}qc\lambda\sqrt{\rhoel(\kk)\rhoel(\kk-\kkph)}\, [\kk\cdot \varepsilon^{(H)}(\kkph)].
\ee

The estimate for $w$ now follows from (\ref {choice:wexpr})
\be
w(\kkph)&>&
\lu^3\int\upd\kk\,\Theta(\mstar-|\kk|)\Theta(\kk\cdot\varepsilon^{(H)}(\kkph))
U^{(H)}(\kkph,-\kk)\cr
&>&
\frac 1{14\mstar}qc\lambda \lu^3\int\upd\kk\,\Theta(\mstar-|\kk|)\Theta(\kk\cdot\varepsilon^{(H)}(\kkph))
\sqrt{\rhoel(\kk)\rhoel(\kk-\kkph)}\, [\kk\cdot \varepsilon^{(H)}(\kkph)]\cr
&=&
\frac 1{14\mstar}qc\lambda \rhoel(0)\lu^3\int\upd\kk\,\Theta(\mstar-|\kkph|)\Theta(\kk\cdot\varepsilon^{(H)}(\kkph))
e^{-\frac 1{4\sigma^2}|\kk|^2-\frac 1{4\sigma^2}|\kk-\kkph|^2}\, [\kk\cdot \varepsilon^{(H)}(\kkph)].\cr
& &
\ee

Introduce a new coordinate system with $\kkph$ in direction 3 and $\varepsilon^{(H)}(\kkph))$ in direction 1.
Then the above becomes
\be
w(\kkph)
&>&
\frac 1{14\mstar}qc\lambda \rhoel(0)e^{-\frac{1}{4\sigma^2}|\kkph|^2}
\lu^3\int_0^\infty\upd k_1\,k_1\int\upd k_2\,\int\upd k_3\,
\Theta(\mstar-|\kk|)\Theta(k_1)
e^{-\frac 1{2\sigma^2}|\kk|^2}e^{\frac 1{2\sigma^2}k_3|\kkph|}\cr
&=&
\frac 1{14\mstar}qc\lambda \rhoel(0)e^{-\frac{1}{4\sigma^2}|\kkph|^2}
\lu^3\int_0^\mstar r^2\upd r\,\int_{-\pi/2}^{\pi/2}\sin(\theta)\upd\theta\int_{-\pi/2}^{\pi/2}\upd\phi\,
\cos(\theta)\cos(\phi)\, e^{-\frac1{2\sigma^2}r^2}e^{\frac 1{2\sigma^2}r\sin\theta}\cr
&=&
\frac {\pi}{14\mstar}qc\lambda \rhoel(0)e^{-\frac{1}{4\sigma^2}|\kkph|^2}
\lu^3\int_0^\mstar r^2\upd r\,e^{-\frac1{2\sigma^2}r^2}\int_{-\pi/2}^{\pi/2}\sin(\theta)\upd\theta
\cos(\theta)\, e^{\frac 1{2\sigma^2}r\sin\theta}\cr
&=&
\frac {\pi}{14\mstar}qc\lambda \rhoel(0)e^{-\frac{1}{4\sigma^2}|\kkph|^2}
\lu^3\int_0^\mstar r^2\upd r\,e^{-\frac1{2\sigma^2}r^2}\int_{0}^{\pi/2}\sin(\theta)\upd\theta
\cos(\theta)\, \cosh(\frac 1{2\sigma^2}r\sin\theta)\cr
&>&
\frac {\pi}{14\mstar}qc\lambda \rhoel(0)e^{-\frac{1}{4\sigma^2}|\kkph|^2}
\lu^3\int_0^\mstar r^2\upd r\,e^{-\frac1{2\sigma^2}r^2}\int_{0}^{\pi/2}\sin(\theta)\upd\theta
\cos(\theta)\cr
&=&
\frac {\pi}{28\mstar}qc\lambda \rhoel(0)e^{-\frac{1}{4\sigma^2}|\kkph|^2}
\lu^3\int_0^\mstar r^2\upd r\,e^{-\frac1{2\sigma^2}r^2}\cr
&=&
\lu^3\pi\sigma^2qc\lambda \rhoel(0)e^{-\frac{1}{4\sigma^2}|\kkph|^2}f\left(\frac{\mstar}{\sigma}\right)
\ee
for $f$ given by
\be
f(u)&=&\frac {1}{28}\int_0^{u} t^2\upd t\,e^{-\frac12 t^2}.
\ee

\bigskip

On the other hand is
\be
\hbar c|\kkph|\Zph(\kkph)
&=&
\hbar c|\kkph|\lu^3\int\upd\kk\,\Theta(\kk\cdot\varepsilon^{(H)}(\kkph))\rhoel(\kk)\cr
&=&
\hbar c|\kkph|\rhoel(0)\lu^3\int\upd\kk\,\Theta(\kk\cdot\varepsilon^{(H)}(\kkph))
\exp\left(-\frac{1}{2\sigma^2}|\kk|^2\right)
\cr
&=&
\hbar c|\kkph|\rhoel(0)\lu^3 2\pi\int_0^\infty r^2\upd r\,e^{-\frac{1}{2\sigma^2}r^2}\cr
&=&
\hbar c|\kkph|\rhoel(0)\lu^3 2\pi\sigma^3\int_0^\infty t^2\upd t\, e^{-\frac12 t^2}\cr
&=&
\hbar c|\kkph|\rhoel(0)\lu^3 2\pi\sigma^3\sqrt{2\pi}.
\ee

\bigskip

One concludes that (\ref {long:lemmacon}) holds for $\epsilon$ given by
\be
\epsilon=\frac{1}{\sqrt{2\pi}}\lambda q f\left(\frac{\mstar}{\sigma}\right).
\ee

\section*{}

\end{document}